\documentstyle[12pt]{article}
\global\parskip 6pt

\normalsize

\let\oldtheequation=\theequation
\def\doteqs#1{\setcounter{equation}{0}
            \def\theequation{{#1}.\oldtheequation}}

\newcounter{sxn}
\def\sx#1{\addtocounter{sxn}{1} \bigskip\medskip \goodbreak
\noindent{\large\bf
\centerline{\thesxn.~~#1}} \nobreak \medskip}
\def\sxn#1{\sx{#1} \doteqs{\thesxn}}

\newcounter{axn}

\def\br{\begin{thebibliography}{abc}}
\def\er{\end{thebibliography}}

\def\be{\begin{equation}}
\def\ee{\end{equation}}
\def\bea{\begin{eqnarray}}
\def\eea{\end{eqnarray}}

\begin{document}
\begin{flushright}
\hfill{SINP-TNP/01-28}\\
\end{flushright}
\vspace*{1cm}
\thispagestyle{empty}
\centerline{\large\bf Further Evidence for the Conformal Structure of a }
\centerline{\large\bf Schwarzschild Black Hole in an Algebraic Approach}
\bigskip
\begin{center}
Kumar S. Gupta\footnote{Email: gupta@theory.saha.ernet.in}\\
{\em Saha Institute of Nuclear Physics\\
1/AF Bidhannagar\\
Calcutta - 700 064, India}\\
\vspace*{.5cm}
Siddhartha Sen\footnote{Email: sen@maths.tcd.ie}\\
{\em School of Mathematics\\
Trinity College\\
Dublin , Ireland}\\
\vspace*{.1cm}
and\\
\vspace*{.1cm}
{\em Department of Theoretical Physics}\\
{\em Indian Association for the Cultivation of Science}\\
{\em Calcutta - 700032, India}\\
\end{center}
\vskip.5cm

\begin{abstract}
We study the excitations of a massive Schwarzschild black hole of mass $M$
resulting from the capture of infalling matter described by a massless scalar 
field. The near-horizon dynamics of this system is governed by
a Hamiltonian which is related to the Virasoro algebra 
and admits a one-parameter family of self-adjoint extensions described by a
parameter $z \in R $. The density of states of the black hole can be 
expressed equivalently in terms of $z$
or $M$, leading to a consistent relation between these two
parameters. The corresponding black hole entropy is obtained 
as $ S = S(0) - \frac{3}{2} {\rm log} S(0) + C$, where $S(0)$ is 
the Bekenstein-Hawking entropy, $C$ is a constant with other subleading
corrections exponentially suppressed. 
The appearance of this precise form for the black hole entropy within our
formalism, which is
expected on general grounds in any conformal field theoretic description, 
provides strong evidence 
for the near-horizon conformal structure in this system.
\end{abstract}
\vspace*{.3cm}
\begin{center}
December 2001
\end{center}
\vspace*{1.0cm}
\baselineskip=18 pt
\newpage
\sxn{Introduction} 

The notion of the existence of a conformal field theory in the near-horizon 
region of a black hole has recently led to interesting developments 
in quantum gravity \cite{strom,carlip,solo,amit}. By imposing
a suitable boundary condition at the horizon, it has been shown that the 
algebra
of surface deformations contains a Virasoro subalgebra. This approach
is based on extension of the idea
of Brown and Henneaux \cite{brown} regarding asymptotic
symmetries of three-dimensional anti-de-Sitter gravity.

In a previous paper \cite{ksg} we proposed a novel algebraic approach 
for analyzing 
the near horizon conformal structure of a Schwarzschild black hole.
Our method uses the time independent modes 
of a scalar field to probe the near-horizon geometry. The dynamics of the 
scalar field
in the near-horizon region is described by a static Klein-Gordon (KG)
operator. This operator, which is the Hamiltonian for the system,
belongs to the representation space of the Virasoro algebra with central
charge $c = 1$.
In the quantum theory, it admits a one-parameter family of self-adjoint
extensions labelled by $e^{i z}$ where $z$ is a real number. For a generic
value of $z$, the system admits an infinite number of bound 
states. When $z$ is positive and satisfies the consistency condition 
$z \sim 0$, the bound states exhibit 
a scaling behaviour in the near-horizon region of the black hole. This
property of the bound states reflects the existence of an underlying 
conformal structure in the near-horizon region. 

In this Letter we provide further evidence for the existence of a
near-horizon conformal structure of a Schwarzschild black hole.
Our analysis is based on the identification of the bound states described
above with the quantum excitations of the black hole resulting from the 
capture of the scalar field probe. These excitations  
will eventually decay through the emission of Hawking radiation. While
the time-independent modes of the scalar field cannot describe the actual
decay process, they do carry information about the static thermal properties
of the system. Within our formalism, any such information must be obtained
from the properties of the associated Hamiltonian. The spectrum of the
Hamiltonian is determined by boundary conditions which are encoded in the
choice of the corresponding domain. Since the self-adjoint parameter $z$
labels the domains of the Hamiltonian, it is therefore directly related
to the boundary conditions. At an operator level, the properties of the 
Hamiltonian are thus determined by the parameter $z$. 
On the other hand, the only physical input in the problem is the mass
$M$ of the black hole which must also play a role in determining
the spectrum. The parameters $z$ and $M$ thus play a
conceptually similar role and they are likely to be related to each other. 
Establishing a relation between $z$ and $M$, which is consistent with the
constraints of our model, is the first step in this approach. 

	The relationship between $z$ and $M$ naturally leads to the 
identification of black hole entropy within this formalism. The density of
states of a black hole is usually a function of the mass $M$ \cite{hooft}.
On the other hand, the density of states in our algebraic formalism has a 
smooth expression in terms of the self-adjoint parameter $z$.
In view of the
relation between $z$ and $M$, the density of states written as a function 
of $z$ can be reexpressed in terms of the variable $M$.
This process leads to the  
identification of the black hole entropy consistent with the constraints of
the model. The entropy so obtained
is given by the Bekenstein-Hawking term together with
a leading logarithmic correction whose coefficient is $- \frac{3}{2}$. The 
other subleading corrections terms are found to be exponentially suppressed.
Such a logarithmic correction term to the Bekenstein-Hawking entropy 
with a $- \frac{3}{2}$ coefficient was first 
found in the quantum geometry formalism \cite{partha} and has subsequently
appeared in several other publications \cite{car1,trg1,das1,danny,partha1}.
In particular, using an exact convergent expansion for the partition of a 
number due to Rademacher \cite{rade}, rather than the asymptotic formula due
to Hardy and Ramanujan \cite{ram}, it was shown in Ref. \cite{danny} 
that within any conformal field theoretic description, the black hole
entropy can be expressed in an exact and convergent series
where the leading logarithmic correction to Bekenstein-Hawking term always  
appears with a universal coefficient of $- \frac{3}{2}$ with the other 
subleading terms exponentially suppressed.
Here we show that the entropy of the Schwarzschild black hole contains 
correction terms precisely of this structure, which provides a strong 
evidence for the underlying near-horizon conformal structure of the system.

	The organization of this paper is as follows. In Section 2 we
briefly describe the near-horizon quantization indicating how the
self-adjoint parameter $z$ appears. Section 3 provides the identification
between $z$ and $M$ leading to the expression of the black hole entropy
within this formalism. We conclude the paper in Section 4 with a discussion
of the main results. 

\sxn{Near-horizon Quantization}

The Klein-Gordon (KG) operator for the time-independent modes of a scalar 
field in the background of a Schwarzschild black hole of mass $M$ 
is given by \cite{trg}
\be
H = - {\frac{d^2}{dx^2}} - {1 \over {4 x^2}},
\ee
where $x = r - 2 M$ is the near horizon coordinate and $r$ is the radial 
variable. In what follows we shall
assume that the black hole is massive, i.e. $M$ is large.
The scalar field $\psi$ in this background obeys the eigenvalue equation
\be
H \psi = {\cal E} \psi,
\ee
where $\psi \in L^2[R^+, dr]$. 
Eqn. (2.1) is an example of an unbounded linear operator on a Hilbert space.
Below we shall first summarize some basic properties of these operators
which would be useful for our analysis.

        Let $T$ be an unbounded differential operator acting on a Hilbert
space ${\cal H}$ and let $(\gamma , \delta )$ denote the inner product
of the elements $\gamma , \delta \in {\cal H}$.
By the Hellinger-Toeplitz theorem \cite{reed}, $T$  has a well defined    
action only on a dense subset $D(T)$ of the Hilbert space  ${\cal H}$.    
$D(T)$ is known as the domain of the operator $T$. Let $D(T^*)$ be the set
of $\phi \in {\cal H}$ for which there is a unique $\eta \in {\cal H}$ with 
$(T \xi , \phi) = (\xi , \eta )~ \forall~ \xi \in D(T)$. For each such
$\phi \in D(T^*)$ we define $T^* \phi = \eta$. $T^*$ is called the adjoint  
of the operator $T$ and $D(T^*)$ is the corresponding domain of the adjoint.

The operator $T$ is called symmetric or Hermitian if $T \subset T^*$,     
i.e. if $D(T) \subset D(T^*)$ and $T \phi = T^* \phi~ \forall~ \phi \in   
D(T)$. Equivalently, $T$ is symmetric iff $(T \phi, \eta) = (\phi, T \eta)
~ \forall ~ \phi, \eta \in D(T)$. The operator $T$ is called self-adjoint
iff $T = T^*$ and $D(T) = D(T^*)$.

We now state the criterion to determine if a symmetric operator $T$ is
self-adjoint. For this purpose let us define the deficiency subspaces 
$K_{\pm} \equiv {\rm Ker}(i \mp T^*)$ and the 
deficiency indices $n_{\pm}(T) \equiv
{\rm dim} [K_{\pm}]$. $T$ is (essentially) self-adjoint iff
$( n_+ , n_- ) = (0,0)$.
$T$ has self-adjoint extensions iff $n_+ = n_-$. There is a one-to-one
correspondence between self-adjoint extensions of $T$ and unitary maps
from $K_+$ into $K_-$. Finally if $n_+ \neq n_-$, then $T$ has no
self-adjoint extensions.

We now return to the discussion of the operator $H$.
On a domain $ D(H) \equiv \{\phi (0) = \phi^{\prime} (0) = 0,~
\phi,~ \phi^{\prime}~  {\rm absolutely~ continuous} \} $, $H$ is a symmetric
operator with deficiency indices (1,1). The corresponding deficiency
subspaces $ K_{\pm} $ are 1-dimensional and are spanned by
\bea
\phi_+ (r) &=& r^{\frac{1}{2}}H^{(1)}_0 (re^{i \frac{ \pi}{4}}),\\
\phi_- (r) &=& r^{\frac{1}{2}}H^{(2)}_0 (re^{-i \frac{ \pi}{4}})  
\eea
respectively, where $ H^{(1)}_0 $ and $ H^{(2)}_0 $ are Hankel functions.
The operator $H$ is not self-adjoint on
$D(H)$ but admits a one-parameter family of self-adjoint extensions. The are
labelled by 
unitary maps from $ K_+ $ into $ K_- $.  The self-adjoint
extensions of $H$ are thus labelled by $e^{i z}$ where $z \in R$. 
Each value of the parameter $z$ defines a domain $D_z(H)$ on which $H$ is 
self-adjoint and thus corresponds to a particular choice of boundary
condition. 

The normalized bound state eigenfunctions and 
eigenvalues of Eqn. (2.1) are given by
\be
\psi_n (x) =  \sqrt{2 E_n x} K_0\left( \sqrt{E_n} x\right)
\ee
and 
\be
{\cal E}_n = -E_n = - {\exp}\left[\frac{\pi}{2} (1 - 8n) {\cot} 
\frac{z}{2}\right]
\ee
respectively, where $n$ is an integer
and $K_{0}$ is the modified Bessel function \cite{narn,trg}. Thus 
for each value of $z$, the operator $H$ admits an infinite number of negative 
energy solutions. In our formalism, these solutions are interpreted as bound
state excitations of the black hole due to the capture of the scalar field.
As is obvious from Eqns. (2.5) and (2.6), different choices of $z$ leads to
inequivalent quantization of the system. The physics of the system is thus
encoded in the choice of the parameter $z$.

The self-adjoint parameter $z$ so far has been kept arbitrary. The
requirement of near-horizon conformal symmetry places important constraint
on $z$ \cite{ksg}. To see this, consider a band-like region
$\Delta = [x_{0} -\delta/\sqrt{E_{0}}, x_{0} + \delta/\sqrt{E_{0}}]$,
where $x_0 \sim \frac{1}{\sqrt{E_0}}$ and 
$\delta \sim 0$ is real and positive. 
For $\Delta $ to belong near the horizon of the black hole, $z$ must be
positive and satisfy the condition $z \sim 0$. When this condition is
fulfilled, at any point $x \in \Delta$ the leading behaviour of
the bound state wavefunction is given by
\be
\psi_n = \sqrt{2 E_n x} \left(A + 2 \pi n\; {\cot} \frac{z}{2}\right),
\ee
where $A = \ln 2 - \gamma$ and $\gamma$ is Euler's constant.
From Eqn. (2.7) we see that the bound state wavefunction exhibits a scaling
behaviour near the black hole horizon which is a reflection of the
underlying conformal structure in the quantum theory. It is important to
note that the consistency condition on $z$ plays a crucial role in obtaining
this result.

\sxn{Density of States and Entropy}

In the analysis presented above, the information about the spectrum of the
Hamiltonian in the Schwarzschild 
background is coded in the parameter $z$. The wavefunctions and 
the energies of the bound states depends smoothly on $z$. Thus, within the
near-horizon region $\Delta $, we propose to identify
\be
{\tilde \rho} (z) \equiv \sum^{\infty}_{n=0} |\psi_n (z)|^2 
\ee
as the density of states for this system written in terms of the variable
$z$. ${\tilde \rho} (z) dz$ 
counts the number of states when the self-adjoint parameter lies between $z$
and $z + dz$. As mentioned before, within the region $\Delta $ 
$z$ is positive and
satisfies the consistency condition $z \sim 0$ \cite{ksg}. From Eqns. (2.6)
and (2.7) we therefore see that the term
with $n=0$ provides the dominant contribution 
to the sum in Eqn. (3.1). The contribution of the terms with 
$n \neq 0$ to the
sum in Eqn. (3.1) is exponentially small for large ${\rm cot} \frac{z}{2}$.
Physically this
implies that the capture of the minimal probe excites only the lowest energy 
state in the near-horizon region of the massive black hole. 
The density of energy states of the black hole in the region $\Delta$ can
therefore be written as
\be
{\tilde \rho} (z)  \approx | \psi_0 |^2 = 
2 A^2 ~ e^{ {{ \pi} \over  4} {\rm cot} {z \over 2}}.
\ee
As mentioned before, within the region $\Delta$, ${\rm cot} {z \over 2}$
is a large and positive number. We thus find that 
the density of states of a massive black hole
is very large in the near-horizon region.

In order to proceed, we shall first provide a physical interpretation of the
self-adjoint parameter $z$ using the Bekenstein-Hawking entropy formula. 
To this end, recall that in our formalism, the capture of the scalar field
probe gives rise to the excitations of the black hole which 
subsequently decay by emitting Hawking radiation. 
A method of deriving density of states and entropy 
for a black hole in a similar physical setting using quantum mechanical
scattering theory
has been suggested by 't Hooft \cite{hooft}. This simple
and robust derivation uses the black hole
mass and the Hawking temperature as the only physical inputs and is
independent of the microscopic details of the system.
The interaction of infalling matter
with the black hole is assumed to be described by Schrodinger's
equation and the relevant emission and absorption cross sections 
are calculated using Fermi's Golden Rule. Finally, 
time reversal invariance (which is equivalent to CPT invariance in
this case) is used to relate the emission and absorption cross sections.
The density of states for a massive black hole of mass $M$ obtained
from this scattering calculation is given by 
\be
\rho (M) = e^{4 \pi M^2 + C^{\prime}} = e^S,
\ee
where $C^{\prime}$ is a constant and $S$ is the black hole entropy. 
It may be noted
that for the purpose of deriving Eqn. (3.3) the infalling matter was
described as particles. However, 
the above derivation of the density of states is independent of the
microscopic details and is valid for a general class of infalling matter. 

We are now ready to provide a physical interpretation of the parameter $z$. 
First note that the density of states calculated in Eqns. (3.2) and (3.3)
correspond to the same physical situation described in terms of different
variables. In our picture, the near horizon dynamics of the scalar field
probe contains information regarding the black hole
background through the self-adjoint parameter $z$. The same information
in the formalism of 't Hooft is contained in the black hole mass $M$. It is
thus meaningful to relate the density of states 
in our framework (cf. Eqn. 3.2) to that given by Eqn. (3.3). If these
expression describe the same physical situation, we are led to the
identification
\be
\frac{ \pi }{4} {\rm cot} {z \over 2} = 4 \pi  M^2.
\ee
Note that the analysis presented here is valid only for massive black holes. 
We have also seen that in the near-horizon region $z$ must be positive and 
obey a consistency condition such that
that ${\rm cot} {z \over 2}$ is large positive number. Thus the relation
between $z$ and $M$ given by Eqn. (3.4) is consistent with the constraints
of our formalism.
We therefore conclude that the self-adjoint parameter $z$ has a physical
interpretation in terms of the mass of the black hole.

As stated above, the density of states of the black hole can be expressed 
either in terms of the variable $z$ or in terms of $M$. In view of the
relation between $z$ and $M$, we can write 
\be
{\tilde \rho} (z) dz \sim |J| \rho (M) dM,
\ee
where $J = \frac{d z}{d M}$ is the Jacobian of the transformation from the
variable $z$ to $M$. When $z \sim 0$, using Eqns. (3.3) and (3.4) we get
\be
{\tilde \rho} (z) dz \sim e^{4 \pi M^2} \frac{1}{M^3} dM 
\sim e^{4 \pi M^2 - \frac{3}{2} {\rm ln} M^2} dM .
\ee
The presence of the logarithmic correction term in the above equation is
thus due to the effect of the Jacobian.

Finally, the entropy for the Schwarzschild black hole obtained from 
Eqn. (3.6) can be written as 
\be
S =  S(0) - \frac{3}{2} {\rm ln} S(0) + C ,
\ee
where $ S(0) = 4 \pi M^2$ is the Bekenstein-Hawking entropy and 
$C$ is a 
constant. Thus the leading correction to the Bekenstein-Hawking entropy is 
provided by the logarithmic term in Eqn. (3.7) with a coefficient of
$- \frac{3}{2} $. The subleading corrections to the entropy coming from the 
$n \neq 0 $ terms of Eqn. (3.1) are 
exponentially suppressed. 
As stated before, this is precisely the structure associated with the
expression of black hole entropy whenever the same is calculated within a
conformal field theoretic formalism \cite{danny}. 
We are thus led to conclude that the expression of the black hole entropy
obtained in our formalism provides a strong indication of
the underlying near horizon-conformal structure present in the system.

\sxn{Conclusion}

In this Letter we have extended the analysis based on the 
algebraic formalism developed in Ref.
\cite{ksg} to provide further evidence for the
near-horizon conformal structure of a Schwarzschild black hole.
Earlier, time
independent modes of a scalar field have been used to probe the near-horizon
geometry of the black hole. The near-horizon dynamics was described by a
Hamiltonian that belongs to the representation space of the Virasoro
algebra,
which for the Schwarzschild background has a central charge $c = 1$
\cite{ksg}.\footnote{It may be noted that the representation space for the 
$c=1$ conformal field theory
obtained in Ref. \cite{ksg} is spanned by tensor densities of weight
$\frac{1}{2}$, i.e. spin $\frac{1}{2}$ \cite{kac}.}
In the quantum theory, this Hamiltonian admits a one-parameter family of
self-adjoint extensions giving rise to bound states in the corresponding
spectrum. When $z$ is positive and satisfies the consistency condition
$z \sim 0$, the bound states exhibit a scaling behaviour in the near-horizon
region of the black hole.

In order to investigate this idea further, we first observe that the 
self-adjoint
parameter $z$ describes the domain of the Hamiltonian and directly
determines the spectrum within this formalism. On the other hand, the black
hole mass $M$ must also play a role in determining the spectrum. 
These two parameters thus play a conceptually similar role and it is 
expected that they will be related. 

The next step of our 
analysis was based on the identification of these bound states with the
excitations of the black hole resulting from the capture of the scalar field
probe. 
These excitation would eventually decay through the emission of
Hawking radiation. This process is described by 
quantum mechanical
scattering theory in terms of the density of states of the black hole
which is a function of the variable $M$ \cite{hooft}.
On the other hand, the
density of states following from our formalism is a smooth
function of $z$. Identifying 
these two expressions of the density of states 
leads to a quantitative relation between $z$ and $M$ which is
consistent with the constraints of the system. Such a relation also
provides a physical interpretation of the self-adjoint parameter in terms of
the mass of the black hole.

The relation between $z$ and $M$ naturally leads to the identification of
black hole entropy within this formalism. The entropy thus obtained 
contains the usual Bekenstein-Hawking term together with 
a leading logarithmic correction which has  
$- \frac{3}{2}$ as the coefficient. Moreover, 
the subleading non-constant corrections are shown to
be exponentially suppressed. It has been observed that 
the expression for the black hole entropy is expected to have precisely this
structure whenever it is calculated within the conformal field 
theoretic formalism \cite{danny} and possibly even for the non-unitary case
\cite{moore}.
Thus the expression that we obtain for the 
black hole entropy provides strong support to the hypothesis 
of an underlying conformal structure in the near-horizon region of the 
Schwarzschild black hole.

Finally we would like to mention that the Hamiltonian appearing in Eqn.
(2.1) can be used to describe the near-horizon dynamics in other black hole
backgrounds with different coefficient of the inverse-square term
\cite{trg}. It is thus plausible that the method developed in this Letter
can be used to analyze the entropy for other black holes as well.


\bibliographystyle{unsrt}

\begin{thebibliography}{99}
\bibitem{strom} A. Strominger, JHEP {\bf 9802} (1998) 009.
\bibitem{carlip} S. Carlip, Phys. Rev. Lett. {\bf 82} (1999) 2828; Class.
Quant. Grav. {\bf 16} (1999) 3327; Nucl. Phys. Proc. Suppl. {\bf 18} (2000) 
10. 
\bibitem{solo} S. N. Solodukhin, Phys. Lettt. {\bf B 454} (1999) 213.
\bibitem{amit} O. Dreyer, A. Ghosh and J. Wisniewski, Class. Quant. Grav.
{\bf 18} (2001) 1929.
\bibitem{brown} J. D. Brown and M. Henneaux, Commun. Math. Phys. {\bf 104}
(1986) 207.
\bibitem{ksg} Danny. Birmingham, Kumar S. Gupta and Siddhartha Sen,
Phys. Lett. {\bf B505} (2001) 191.
\bibitem{hooft} G 't Hooft, Int. Jour. Mod. Phys {\bf A11} (1996) 4623;
Talk at the International School of Subnuclear Physics, Erice, 1999, 
hep-th/0003004.
\bibitem{partha} Romesh K. Kaul and Parthasarathi Majumdar, Phys. Rev. Lett.
{\bf 84} (2000) 5255.
\bibitem{car1} S. Carlip, Class. Quant. Grav. {\bf 17} (2000) 4175.
\bibitem{trg1} T. R. Govindarajan, Romesh K. Kaul and V. Suneeta, 
Class. Quant. Grav. {\bf 18} (2001) 2877.
\bibitem{das1} Saurya Das, Romesh K. Kaul and 
Parthasarathi Majumdar, Phys. Rev. {\bf D 63} (2001) 044019.
\bibitem{danny} Danny Birmingham and Siddhartha Sen, Phys. Rev. {\bf D 63},
(2001) 04750.
\bibitem{partha1} Saurya Das, Parthasarathi Majumdar and Rajat K. Bhaduri,
hep-th/0111001.
\bibitem{rade} H. Rademacher, {\it Topics in Algebraic Number Theory}, 
Springer-Verlag, Berlin, 1973.
\bibitem{ram} G. H. Hardy and S. Ramanujan, Proc. Lond. Math. Soc. {\bf 2},
(1918) 75.
\bibitem{trg} T. R. Govindarajan, V. Suneeta and S. Vaidya,
Nucl. Phys. {\bf B583} (2000) 291.
\bibitem{reed} M. Reed and B. Simon, {\it Methods of Modern Mathematical 
Physics}, volume 1, Academic Press, New York, 1972; volume 2, Academic   
Press, New York, 1975. 
\bibitem{narn} H. Narnhofer, Acta Physica Austriaca {\bf 40}, 306 (1974).
\bibitem{kac} V. G. Kac and A. K. Raina, {\it Bombay Lectures on Highest 
Weight Representations of Infinite Dimensional Lie Algebras}, World 
Scientific, Singapore, 1987.
\bibitem{moore} J. A. Harvey, D. Kutasov, E. J. Martinec and G. Moore,
hep-th/0111154.
\end{thebibliography}

\end{document}